\documentclass[11pt,twoside]{article}
\pdfoutput=1 
\usepackage{asp2010}
\usepackage{float}
\usepackage[font={small,it}]{caption}

\resetcounters

\begin{document}

\title{Contribution of the disks to the SFR in the local Universe using Integral Field Spectroscopy from CALIFA}
\author{Cristina Catal\'an-Torrecilla$^1$,Armando Gil de Paz$^1$, \'Africa Castillo-Morales$^1$, Jorge Iglesias-P\'aramo$^2$$^,$$^3$, and CALIFA Collaboration}
\affil{$^1$Departamento de Astrof\'isica y CC. de la Atm\'osfera, Universidad Complutense de Madrid, E-28040, Madrid, Spain}
\affil{$^2$Instituto de Astrof\'isica de Andaluc\'ia-CSIC, Granada, Spain} 
\affil{$^3$Centro Astron\'omico Hispano Alem\'an, Almer\'ia, Spain}

\begin{abstract}
The Calar Alto Legacy Integral Field Area survey (CALIFA survey) is providing Integral Field Spectroscopy (IFS) data in the entire optical
window for a diameter-limited sample of 600 objects in the Local Universe (0.005$<$z$<$0.03). One of the main goals of this survey is to
explore the spatial distribution of the star formation in nearby galaxies free from the limitations associated to either UV (dust attenuation) or narrow-band H$\alpha$ imaging (underlying H$\beta$ absorption). These are limitations that have prevented (until now) carrying out
a detailed study of the evolution of the SFR by components (nuclei, bulges, disks), even locally. This kind of studies are key, for 
example, for understanding how galaxies really evolve from the Blue Cloud to the Red Sequence. 
We will first discuss in detail the validity of the assumption that the SFR given by the extincion-corrected H$\alpha$ is a
good measure of the total SFR by means of cross-comparing this with other SFR estimators, namely the integrated UV+22$\mu$m, UV+TIR, H$\alpha_{\rm{obs}}$+22$\mu$m, or H$\alpha_{\rm{obs}}$+TIR. Only once these effects are properly accounted for we can obtain preliminary results from the spatially-resolved analysis of the contribution of disks to the total SFR in the Local Universe, as a local benchmark for future studies of disks at high redshift.
Our analysis shows that at least in the Local Universe the H$\alpha$ luminosity derived from observations of the CALIFA IFS survey can be used to trace the SFR and that the disk to total (disk + bulge) SFR ratio is on average $\sim$88 $\%$. 
\end{abstract}

\section{Introduction}

The measurement of the star formation rate (SFR) is a crucial parameter to understand the birth and evolution of the galaxies \citep{Kennicutt_98} as it provides information on the amount of the gas in galaxies and how efficient is the formation of stars inside them, which depends strongly on the conditions of the interstellar medium in which they are formed.
Until now, the study of the evolution of the SFR has focussed on the analysis of the integrated SFR in galaxies, with little attention being paid on where in galaxies (nuclei, bulges, disks) SFR takes places and how the SFR in each of these components evolves separately with redshift (see \citet{Freundlich} for a notorius exception in this regard). 
In this respect, a proper calibration of the SFR tracers, where spatially resolved studies can be carried out, is essential. In that way we can compare how the SF of these different spatial components behaves in different wavelengths ranges and/or redshifts. 
In this work, we take advantage of the development of the Integral Field Spectroscopy (IFS) technique using a large well-characterized sample of nearby galaxies from the CALIFA survey that spans the entire color-magnitude diagram \citep[see][]{Sanchez_2012}. This will allow properly determining H$\alpha$ and H$\beta$ fluxes to obtain precise Balmer-decrement measurements to derive H$\alpha$ luminosities corrected for attenuation and the corresponding SFRs.
However, it is critical to first determine that, in a statistically large sample of galaxies like CALIFA, no significant fraction of the SFR is being missed when using the extinction-corrected H$\alpha$ luminosity as SFR estimator. This requires of a combined analysis of this estimator with other estimators of the star formation rate including the continuum UV emission, the total infrared luminosity or monochromatic infrared emission. The estimators considered here come in two types: simple and hybrid SFR recipes \citep[see][for a recent compilation]{Calzetti_2012}.
In this work, we derive integrated extinction-corrected H$\alpha$-based SFRs from the analysis of CALIFA IFS data and compare them with measurements from other SFR tracers. Finally, we analyze the contribution of disks to the total SFR and the variation of this ratio with the morphological type and with the presence of a bar in these galaxies.

\section{Analysis}

This work makes use of 229 CALIFA galaxies that have been observed with the integral-field spectrophotometer PMAS/PPAK, mounted on the Calar Alto 3.5 m telescope, until May 2013. The spectra cover the optical wavelength range ($\lambda$$\lambda$\,3700-7000\,\AA\AA) with a spectral resolution of 6.0\AA (FWHM).

\subsection{Integrated optical spectra}	

For each galaxy we generate an integrated spectrum within the largest elliptical aperture possible in the field of view of our CALIFA datacubes. This aperture matches the axial ratio and position angle of the galaxy D25 ellipse as listed in NED and has a major axis radius of 36 arcsec. 

In order to minimize systematics associated to the stellar continuum subtraction at low-S/N regimes we decided to first spatially integrate the datacube within these apertures and later carry out the necessary corrections to derived total extinction-corrected H$\alpha$ luminosities. The first of these corrections is to subtract the stellar continuum underlying the H$\beta$ and H$\alpha$ spectral features. This is done by means of adjusting a combination of two SSP evolutionary synthesis models of \citet{Vazdekis_2010} to the spectrum obtained for each aperture. Then, we compute the resulting H$\beta$ and H$\alpha$ line emission fluxes, by fitting gaussian functions to the residuals. Once the fluxes from both emission lines are computed we correct the H$\alpha$ flux for dust-attenuation assuming an intrinsic Balmer ratio of 2.86.

\subsection{Bulge-Disk Decomposition}

In order to obtain the SFR in the spheroidal component and in the disk separately, we perform one-dimensional photometric bulge-disk decomposition fitting the light profile of each galaxy using SDSS r-band data available for the whole sample. Once the decomposition is done and the aperture for the bulge is defined (that one at which bulge and disk show similar optical surface brightness), we use the same method as described in the previous subsection to do the analysis of the integrated spectrum obtained for each component.

\section{Results}

\subsection{SFRs tracers comparison: single-band and hybrids tracers}

In this section we describe the results from the analysis of the integrated SFR in our galaxies. We obtained that there is a strong correlation between the hybrids tracers and the extintion-corrected H$\alpha$ tracer (see Figure 1). The hybrids tracers combine luminosities measured directly (observed UV and H$\alpha$) with that of the light emitted by dust after being heated by young massive stars. We show that the extintion-corrected H$\alpha$ tracer is able to recover the entire energy budget from recently-formed stars.
It is worth noting that we have obtained similar results in the case of the single-band tracers, although the hybrids tracers yield a tighter relation than single-wavelength calibrators.

\begin{figure}[h]
\includegraphics[width=65mm]{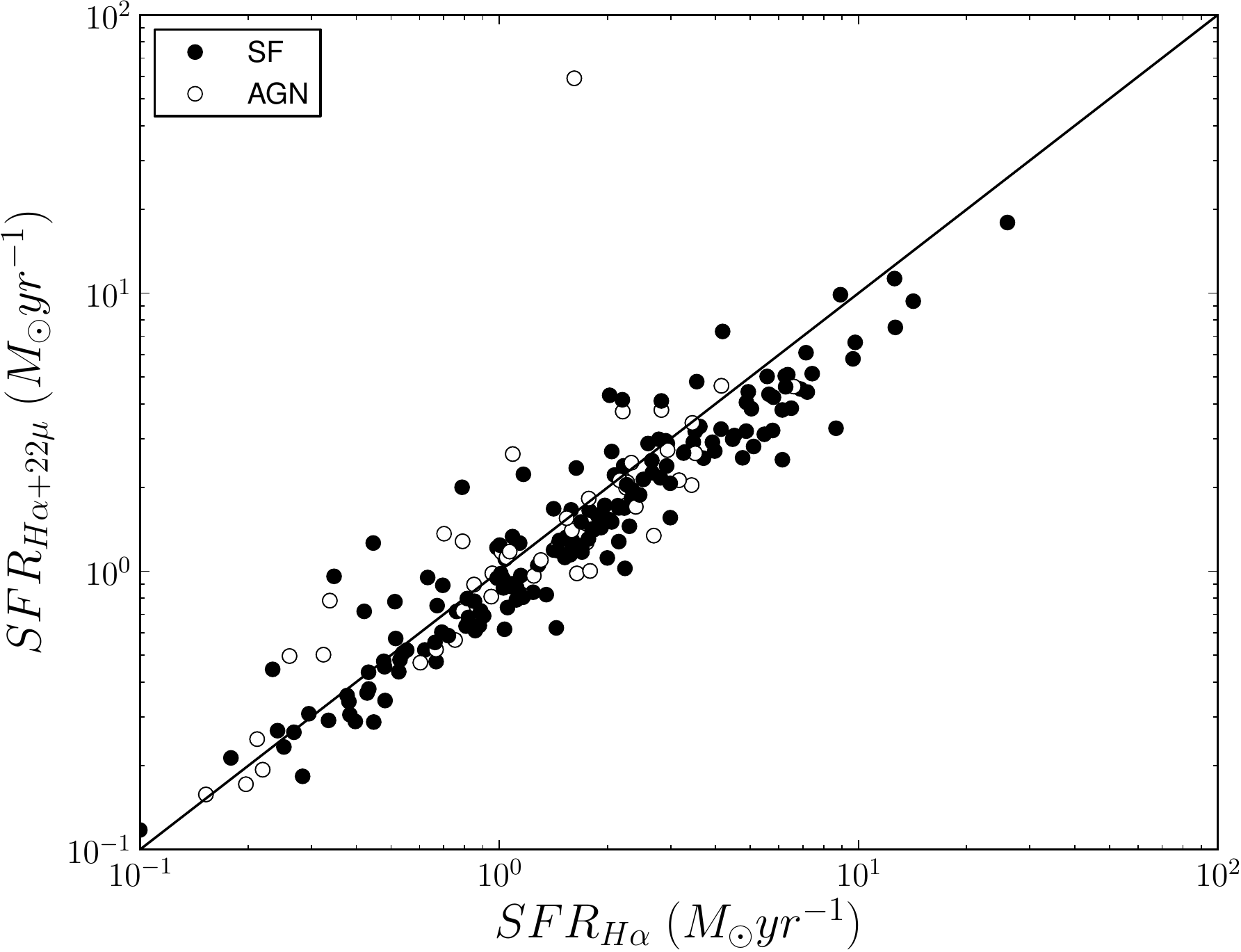}
\hspace{\fill}
\includegraphics[width=65mm]{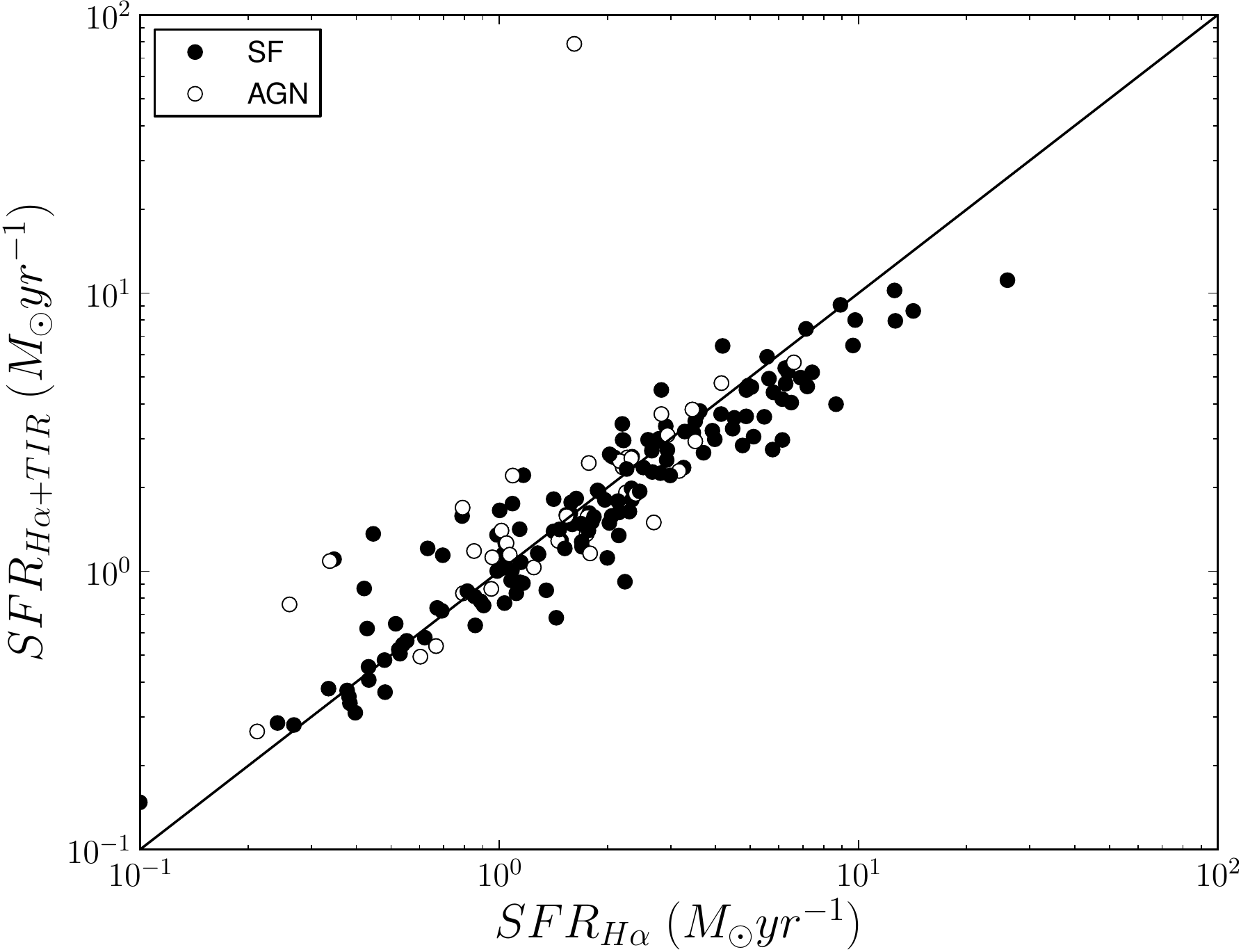}
\hspace{\fill}
\includegraphics[width=65mm]{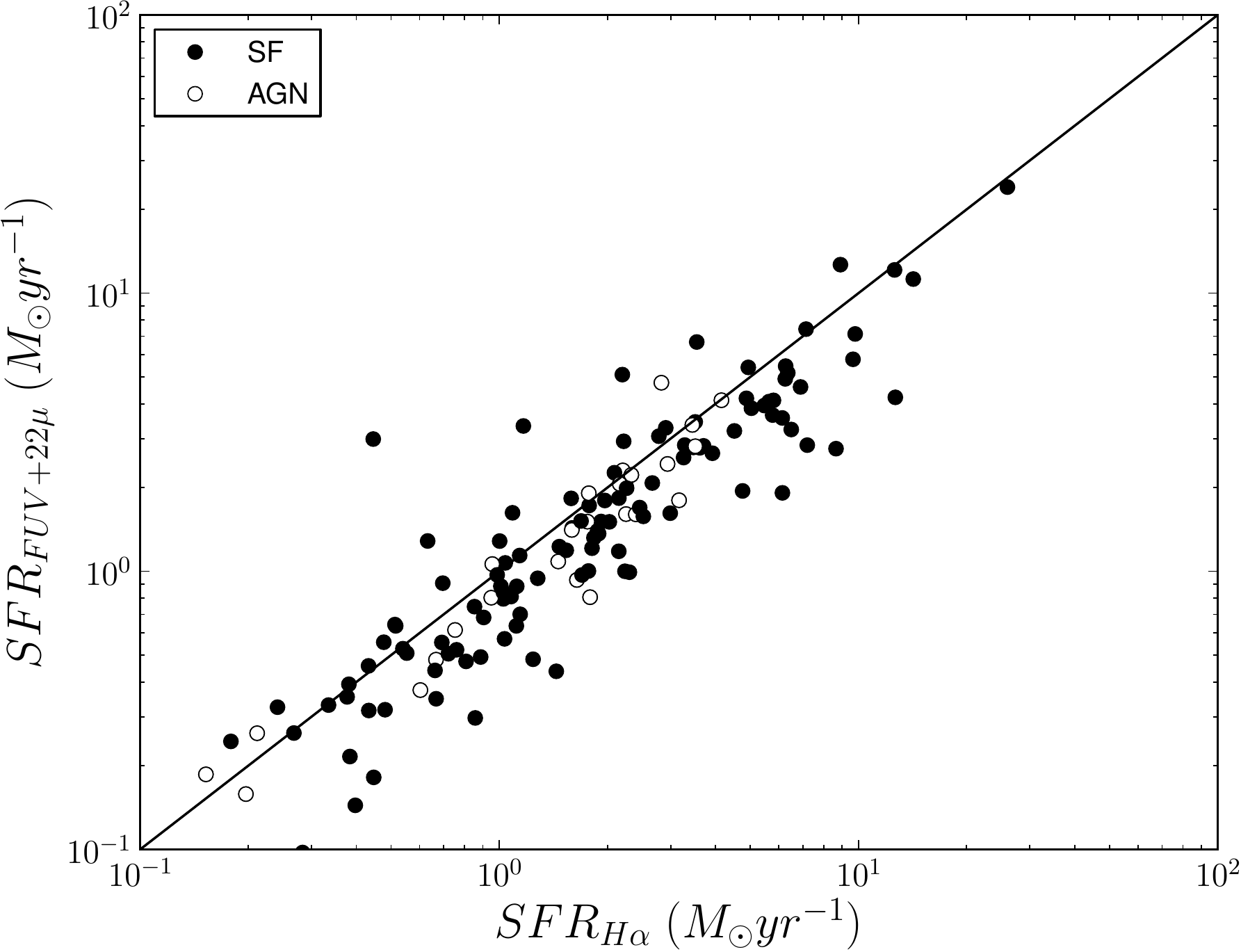}
\hspace{\fill}
\includegraphics[width=65mm]{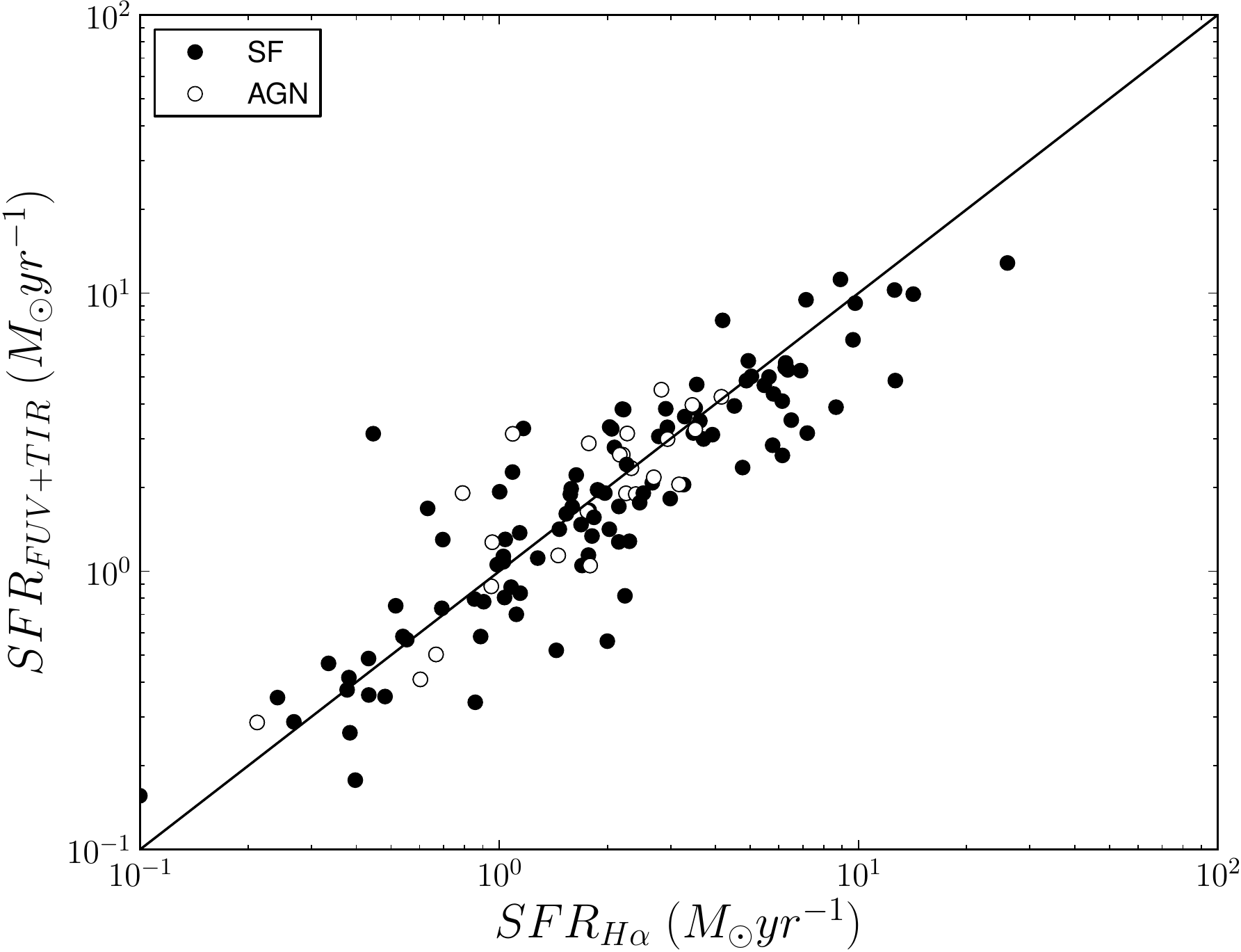}
\caption{Comparison of the different SFRs hybrids tracers plotted against the extincion-corrected H$\alpha$ estimator. Filled circles represent galaxies whose line emission is dominated by star formation and open circles represent galaxies that have been identified as having an AGN. }
\end{figure}

\subsection{Contribution of the disks to the total SFR}

We present preliminary results of the contribution of the disk to the total SFR in each galaxy. We found that the percentage of this ratio, $SFR_{disk}$/$SFR_{total}$, is on average $\sim$88 $\%$. In the case of the spiral galaxies, this ratio oscillates between 0.8 and 1. We do not find significant differences with morphological type; only in the case of the lenticular galaxies we derive somewhat lower values for $SFR_{disk}$/$SFR_{total}$ (see Figure 2).
Our results also point towards the idea that this ratio appears to be the same for barred and unbarred galaxies. Elliptical galaxies have been excluded from this analysis.

\noindent
\begin{minipage}{130mm}% adapt widths of minipages to your needs
\begin{figure}[H]
\centering
\includegraphics[width=80mm]{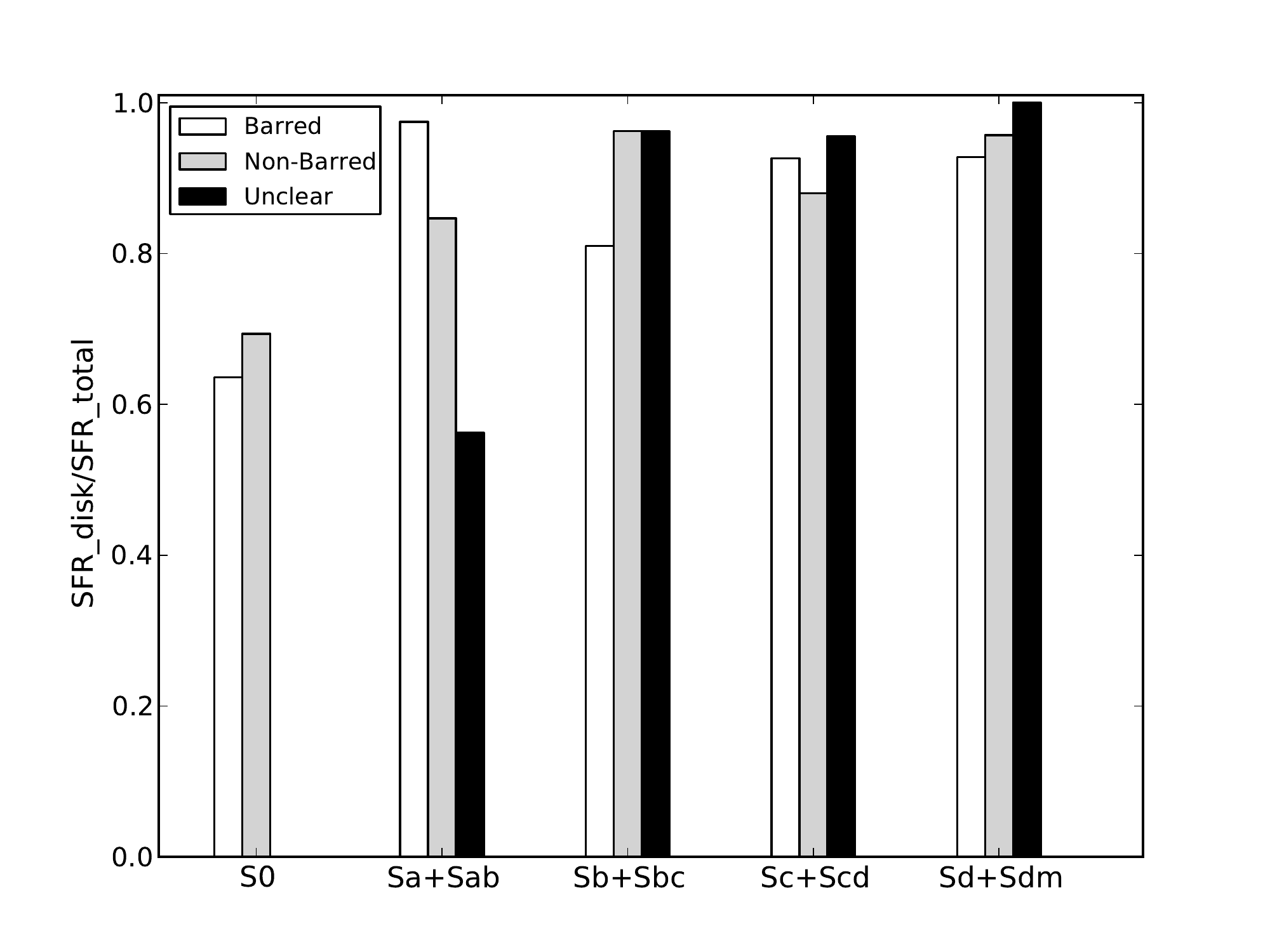}
\caption{Frecuency histogram with the distribution of the $SFR_{disk}$/$SFR_{total}$ as a function of morphological type. White, grey and black columns represent barred, unbarred galaxies and galaxies in which the presence of the bar is not clear, respectively.}
\end{figure}
\end{minipage}%

\section{Conclusions}
In this work, we have first examined the behaviour of the SFR tracers compared with extintion-corrected H$\alpha$ tracer, showing a tight correlation between them, specially when compared to the hybrids ones. We conclude that the H$\alpha$ corrected SFR is an excellent tracer of SF, which takes into account dust attenuation at least in the local Universe and with the selection criteria of our sample. The CALIFA IFS data provides a new capability to study spatially resolved star formation with high spectral resolution H$\alpha$ and H$\beta$. 

We have presented preliminary results of the ratio of the disk to total (disk + bulge) SFR ratio. Our analysis shows that this ratio is on average $\sim$88 $\%$ and it does not depend on the morphological type, but there may be a slightly lower in lenticulars. Our results also point towards the idea that this ratio appears to be rather similar for barred and unbarred galaxies.

\acknowledgements 
This study makes uses of the data provided by the Calar Alto Legacy Integral Field Area (CALIFA) survey (http://califa.caha.es/).

%% lo quito yo \bibliography{editor}

\bibliography{catalanc}

\end{document}